\documentclass[journal]{rmaa}

\usepackage{paralist}

\def\FeH{[Fe/H]}
\usepackage{psfrag,color}

\usepackage[latin1]{inputenc}
\usepackage[T1]{fontenc}
\usepackage{rotating}

\title{A vindication of the RR Lyrae Fourier light curve decomposition for the calculation of  metallicity and distance in globular clusters } 

\author{
  A. Arellano Ferro,\altaffilmark{1}
}

\altaffiltext{1}{Instituto de Astronom\'ia, Universidad Nacional Aut\'onoma de M\'exico, Ciudad Universitaria, C.P. 04510, M\'exico.}

\shortauthor{A. Arellano Ferro}
\shorttitle{Globular clusters Horizontal Branch luminosity}

\abstract{We report the mean metallicity and absolute magnitude of RR Lyrae stars in a sample of 37 globular clusters, calculated via the Fourier decomposition of their light curves and $ad~hoc$ semi-empirical calibrations, in an unprecedented homogeneous approach. This enabled a new discussion of the metallicity dependence of the horizontal branch (HB) luminosity, as a fundamental distance indicator.
The calibration for the RRab and RRc stars should be treated separately. For the RRab the dispersion is larger and  non-linear. For the RRc stars the correlation is less steep, very tight and linear. The relevance of  the HB structural parameter $\mathcal L$, is highlighted and offer a non-linear calibration of the form  $M_V$([Fe/H],$\mathcal L$). Excellent agreement is found between values of [Fe/H] and $M_V$ from the light curve decomposition with spectroscopic values and distances obtained via Gaia-DR3 and HST
The variables census in 35 clusters includes 326 stars found by our program.}

\resumen{Reportamos valores medios de la metalicidad y magnitud absoluta de estrellas RR Lyrae en   37 c\'umulos globulares (CGs), calculados homogeneamente por medio de la descomposici\'on de Fourier de las curvas de luz y de calibraciones $ad~ hoc$ semi emp\'iricas. Lo anterior permiti\'o un nuevo an\'alisis de la luminosidad de la rama horizontal (RH) y su dependencia de la metalicidad, 
como indicador fundamental de distancia. La calibraci\'on para estrellas RRab y RRc debe tratarse por separado. Para las RRab no es lineal y presenta mayor dispersi\'on. Para las RRc la correlaci\'on es lineal y estrecha. 
La relevancia del par\'ametro de estructura de la RH, $\mathcal L$ para las RRab, es evidente. Ofrecemos una calibraci\'on de la forma $M_V$([Fe/H],$\mathcal L$). Los valores de [Fe/H] y $M_V$ comparan muy bien con valores espectrosc\'opicos y determinaciones de distancia obtenidas con datos de $Gaia$-eDR3 and $HST$. El censo de variables en 35 c\'umulos incluye 326 descubiertas por nuestro programa. }

\addkeyword{Globular clusters: general} 
\addkeyword{Stars: horizontal branch}
\addkeyword{Stars: distances}

\addkeyword{Stars: fundamental parameters}

\addkeyword{Stars: variables: RR Lyrae}

\begin{document}
\maketitle

\section{A brief panorama of the $M_V$-[{F\MakeLowercase{e}/H}] relation}
\label{sec:FeHMv}

The relevance of RR Lyrae (RRL) stars as distance indicators has been well known since
the early XXth century. \citet{Shapley1917}  recognized that 
"The median magnitude of the
short-period variables [RR Lyrae stars] apparently has a rigorously constant value in
each globular cluster" and "The observed differences
in the mean values then become sensitive criteria of distance,
and the relative parallaxes of these remote systems can be known with
an accuracy...", a fact that was used later by Shapley himself to describe the
Galactic distribution of globular clusters \citep{Shapley1918}.
This apparently constant value of the mean magnitude of the RRL can now be
interpreted
as the luminosity level of the horizontal branch (HB) being constant in all
globular clusters. The fact that this is not exactly the case, but instead that
metallicity plays a role in determining the luminosity level of the HB, 
has been demonstrated from theoretical and observational grounds. 
 \citet{Sandage1981,LeeDemZinn1990,Sandage1990}
  provided a calibrations of the $M_V$-[Fe/H] relation.
    \citet{LeeDemZinn1990} discuss its dependence on 
  helium abundance. Other empirical calibrations followed in the works of \citet{Walker1992}, Carney et al. (1992), Sandage(1993) and \citet{Benedict2011}.
Complete summaries on the
calibration of the $M_V$-[Fe/H] relation can be found in the works
of \citet{Chaboyer1999}, \citet{CacciariClement2003} and \citet{SandTamm2006}. 

While the relation has been considered to be linear in most  empirical works, a non-linear
nature is advocated by theoretical approaches, e.g.  \citet{Cassisi1999}
and \citet{Vandenberg2000}. A linear relation of the form $M_V$= a+ b[Fe/H] has been broadly accepted in the literature and the slope, resulting from a variety of independent calibrations, ranges  a wide span (0.13 - 0.30) as different strategies have been adopted, mainly towards the calculation of $M_V$. The relevance of the slope and zero point of this relation on the relative and absolute ages of the globular clusters has been amply discussed by \citet{Caboyer1996,Chaboyer1998} and \citet{Demarque2000}.
That a linear relation may be an over simplification  \citep{CatelanSmith2015} becomes very clear from the theoretical analysis of \citet{Demarque2000}, that clearly demonstrates the role of the HB structure and that the slope itself is a function of metallicity. The higher complexity of the HB luminosity and metallicity interconnection has however defy clear empirical demonstrations and calibrations, for which, a very extensive and homogeneous endeavor is required.

Our approach to the determination of mean $M_V$ and [Fe/H] of RRL stars, has been the Fourier light curve decomposition of both the fundamental mode and first overtone pulsators RRab and RRc respectively. This followed by the employment of solidly established semi-empirical calibrations between the Fourier parameters and the physical quantities. As early as 2002, our group started studying individual clusters in great detail from CCD time-series imaging through the Johnson-Kron-Cousins \emph{VI} bands. Each cluster in the sample has been the subject of a dedicated study, and the discussions
include several key aspects of the nature of the globular clusters, such as distinction of likely cluster members from field stars, structure of the colour-magnitude diagram (CMD), pulsating mode distribution on the HB, and theoretical approaches to the mass loss events in the Red Giant branch and the subsequent mass distribution at the stage of core He-burning (zero age horizontal branch or ZAHB) and post ZAHB evolution.

In 2017, \citet{Are2017} (hereinafter ABG17) summarized the results
of a 15-year old program dedicated to study the variable star populations in globular clusters.
The program was mainly aimed to the determination of the mean distance and metallicity of the clusters in an homogeneous way, via the Fourier decomposition of the light curves of their RRL stars and the use of well tested semi-empirical calibrations of the Fourier parameters in terms of luminosity and [Fe/H]. The program is based on the Johnson-Kron-Cousins \emph{VI} CCD time-series imaging, and their subsequent scrutiny  via the difference images analysis (DIA), that produces accurate photometry even in the crowded central regions of the globular clusters. In the process numerous variables of virtually all types typically present in globular clusters were discovered, of which ABG17 gave a detailed account.

Presently, five years after ABG17 paper, our group has systematically continued enlarging  the sample of studied clusters which has increased from 25 to 34.
This should enable a better sustained discussion of the $M_V$-[Fe/H] relation, i.e. the metallicity dependence of the luminosity of the horizontal branch (HB), and the influence of the
cluster Oostherhoff type and the HB structure \citep{Demarque2000}.
In the present paper we update our discussion of the nature and calibration of the  $M_V$-[Fe/H] relation, and introduce the role of the HB structure parameter $\mathcal L$ which is shown to be of an obvious relevance. We also perform a census of variable stars per cluster per variable type and reinforce the resulting cluster distance scale from the Fourier approach via the comparison with independent distances recently obtained from $Gaia$ and HST accurate data.

We shall mention at this point that the parameters listed in the tables below for specific clusters and number of variables, may occasionally differ slightly from the equivalent tables in ABG17, as a result of a critical evaluation of the original samples. The present tables supersede the previous ones.

\section{Observations and image reductions}
\label{sec:ObserRed}

\subsection{Observations}

The observations involved in the program have been obtained in several observatories and telescopes in the 0.8-2.15 m range. The majority of the observations have been performed with the 2.0m
Himalayan Chandra Telescope (HCT) of the Indian Astronomical Observatory
(IAO), Hanle, India. We
have also used the 0.84m of San Pedro M\'artir Observatory (SPM) Mexico, the 2.15-m telescope of the Complejo Astron\'omico El Leoncito
(CASLEO), San~Juan, and the 1.52m telescope of Bosque Alegre of the C\'ordoba Observatory, Argentina, the Danish 1.54 m telescope
at La Silla, Chile, the SWOPE 1.0 m telescope of Las Campanas Observatory, Chile, the LCOGT 1 m telescopes network at the South African
Astronomical Observatory (SAAO) in Sutherland, South
Africa, at the Side Spring Observatory (SSO) in New South Wales, Australia, and at
Cerro Tololo Inter-American Observatory (CTIO), Chile.

\subsection{Transformation to the Standard System}
\label{Tranformation}

All observations have been transformed from the instrumental to the standard  
 Johnson-Kron-Cousins photometric system \citep{Landolt1992}
\emph{VI}, using local standard stars in the fields of the target clusters. These standard stars have been taken from the extensive collection of
\citet{Stetson2000}\footnote{%
 \texttt{http://www3.cadc-ccda.hia-iha.nrc-cnrc.gc.ca/\\
community/STETSON/standards}}.
Typically between 30 and 200 standard stars were available per globular cluster.

\subsection{Difference Image Analysis}
\label{DIA}

All the image photometric treatment has been performed using the Difference Image Analysis using the \emph{DanDIA} pipeline \citep{Bramich2008,Bramich2013,Bramich2015}.

\begin{table*}
\tiny
\begin{center}
\caption{Mean values of [F\MakeLowercase{e}/H], given in three different scales, and $M_V$ from a homogeneous
Fourier decomposition of
the light curves of RRL cluster members.$^1$}
\label{MV_FEH:tab}

\begin{turn}{90}
\begin{tabular}{lc|ccccc|ccccc|ccc}

\hline
GC &Oo& [Fe/H]$_{\rm ZW}$ & [Fe/H]$_{\rm UV}$& [Fe/H]$_{\rm N}$&$M_V$& N& [Fe/H]$_{\rm ZW}$ &[Fe/H]$_{\rm UV}$ &[Fe/H]$_{\rm N}$ & $M_V$ & N & Ref.&$E(B-V)$&$\mathcal L$\\
\hline
NGC (M) & &&& RRab  & &&&& RRc& & & & &\\
\hline
1261     &I &-1.48$\pm$0.05&-1.38&-1.27&0.59$\pm$0.04 &6&-1.51$\pm$0.13 &-1.38 &-1.41&0.55$\pm$0.02  &4&25&0.01& $-0.71$\\

1851      &I &-1.44$\pm$0.10&-1.33&-1.18&0.54$\pm$0.03 &10&-1.40$\pm$0.13&-1.28 &-1.28 & 0.59$\pm$0.02& 5&23&0.02&--0.36\\

3201      &I &-1.49$\pm$0.10&-1.39&-1.29&0.60$\pm$0.04 & 19&-1.47$\pm$0.08&-1.37 &-1.36 & 0.58$\pm$0.01 &2&3&diff.&+0.08\\

4147      &I &--&--&--&-- & -- & -1.72$\pm$0.26&-1.68&-1.66 &0.57$\pm$0.05&6 &  4&0.01&+0.55\\

5272 (M3)  &I &-1.56$\pm$0.16&-1.46&-1.46&0.59$\pm$0.05 &59&-1.65$\pm$0.14&-1.57&-1.56 &0.56$\pm$0.06&23 &24&0.01&+0.08\\

5904 (M5)  &I &-1.44$\pm$0.09&-1.33&-1.19&0.57$\pm$0.08  &35&-1.49$\pm$0.11&-1.39&-1.38 &0.58$\pm$0.03&22 &19&0.03&+0.31\\

6171 (M107)&I &-1.33$\pm$0.12&-1.22&-0.98&0.62$\pm$0.04  & 6 &-1.02$\pm$0.18&-0.90&-0.88 &0.59$\pm$0.03&4 &22&0.33&--0.73\\

6229      &I &-1.42$\pm$0.07&-1.32&-1.13&0.61$\pm$0.06  & 12&-1.45$\pm$0.19&-1.32&-1.58&0.53$\pm$0.10&8&20 &0.01&+0.24\\

6266$^6$ (M62)    &I &-1.31$\pm$0.11&-1.64&--&0.63$\pm$0.03  & 12&-1.45$\pm$0.19&-1.32&-1.58&0.51$\pm$0.03&8&3 &0.47&+0.55\\

6362 &I &-1.25$\pm$0.06&-1.13&-0.83&0.62$\pm$0.01&2&-1.21$\pm$0.15&-1.09&-1.10 &0.59$\pm$0.05&6&27&0.06&--0.58 \\

6366      &I &-0.84&-0.77&-0.31&0.71  & 1 &-- &-- &--&-- & &11$^2$&0.80& --0.97\\

6401      &I &-1.36$\pm$0.09&-1.24&-1.04&0.60$\pm$0.07  &19 &-1.27$\pm$0.23&-1.09 &-1.16 &0.58$\pm$0.03&9& 21&diff&+0.13\\

6712   &I &-1.25$\pm$0.06 &-1.13&-0.82&0.55$\pm$0.03&6 &-1.10$\pm$0.04&-0.95& -0.96&0.57$\pm$0.18 &3&30&0.35&--0.62\\

6934   &I &-1.56$\pm$0.14 &-1.48&-1.49&0.58$\pm$0.05&15 &-1.53$\pm$0.12&-1.41& -1.50&0.59$\pm$0.03 &5&26&0.10&+0.25\\

6981 (M72) &I &-1.48$\pm$0.11&-1.37&-1.28&0.63$\pm$0.02  &12 &-1.66$\pm$0.08&-1.60&-1.55 &0.57$\pm$0.04& 4&14&0.06&+0.14\\

7006  &I &-1.51$\pm$0.13&-1.40&-1.36&0.61$\pm$0.03  &31 &-1.53&-1.44&-1.43&0.55& 1&33&0.08&--0.28\\

Pal13  & I&-1.64$\pm$0.15 &-1.56 &-1.67 & 0.65$\pm$0.05 & 4 &--&--& --&-- & -- &28&0.10 &--0.30\\

\hline

288       & II&-1.64&-1.58&-1.42&  0.38& 1&-1.59 &-1.52 &-1.54 &0.58& 1 &1&0.03&+0.98\\

1904 (M79) & II&-1.63$\pm$0.14&-1.55&-1.47&  0.41$\pm$0.05& 5 &-1.71 &-1.66 &-1.69&0.58 & 1 &2&diff&+0.74\\

4590 (M68) & II&-2.07$\pm$0.09$^3$&-2.21&-2.01&  0.49$\pm$0.07& 5 &-2.09$\pm$0.03&-2.24&-2.23 &0.53$\pm$0.01&15 &5&0.05&+0.17 \\

5024 (M53) & II&-1.94$\pm$0.06$^3$&--2.00&-1.68&  0.45$\pm$0.05 &18&-1.84$\pm$0.13&-1.85&-1.85 &0.52$\pm$0.06& 3 &6 &0.02&+0.81\\

5053      & II&-2.05$\pm$0.14$^3$&-2.18&-2.07&  0.46$\pm$0.08 & 3&-2.00$\pm$0.18&-2.05&-2.06 &0.55$\pm$0.05& 4&7&0.18&+0.50\\

5286$^6$ &II &-1.68$\pm$0.15&-1.64&--&0.52$\pm$0.04 &59&-1.71$\pm$0.23&-1.68&--&0.57$\pm$0.04&23 &3&0.24&+0.80\\

5466      & II&-2.04$\pm$0.14$^3$&-2.16&-2.01&  0.44$\pm$0.09 & 8 &-1.90$\pm$0.21 &-1.89&-1.96&0.53$\pm$0.06 & 5 &8 &0.00&+0.58 \\

6205 (M13)  & II&-1.60 &-1.54 &-1.00 & 0.38  & 1 &-1.70$\pm$0.20  &-1.63&-1.71 &0.59$\pm$0.05 & 3 &29 &0.02&+0.97\\

6254 (M10)  & II?&-- &-- &--& --  & -- &-1.59&-1.52&-1.52 &0.52 & 1 &32 &0.25&+1.00\\

6333 (M9)  & II&-1.91$\pm$0.13$^3$&-1.96&-1.72&  0.47$\pm$0.04& 7 &-1.71$\pm$0.23 &-1.66&-1.66 &0.55$\pm$0.04& 6 & 9&diff&+0.87\\

6341 (M92) & II&-2.12$\pm$0.18$^3$&-2.16$^5$&-2.26&   0.45$\pm$0.03& 9&-2.01$\pm$0.11&-2.11&-2.17 &0.53$\pm$0.06& 3 &10 &0.02&+0.91\\

6809$^6$  &II &-1.61$\pm$0.20&-1.55&--&0.53$\pm$0.09 &59&--&--&--&--&--&3&0.08&+0.87\\

7078 (M15) & II&-2.22$\pm$0.19$^3$&-2.46&-2.65&  0.51$\pm$0.04 &9&-2.10$\pm$0.07&-2.24&-2.27 &0.52$\pm$0.03 & 8 &15&0.08&+0.67 \\

7089 (M2)  & II&-1.60$\pm$0.18&-1.51&-1.25 & 0.53$\pm$0.13& 10 &-1.76$\pm$0.16&-1.73&-1.76 &0.51$\pm$0.08& 2&16 &0.06&+0.38$^4$\\

7099 (M30) & II&-2.07$\pm$0.05$^3$&-2.21&-1.88&  0.40$\pm$0.04& 3&-2.03  &-2.14&-2.07 &0.54& 1 &17&0.03&+0.89\\

7492      & II&-1.68&-1.63&-0.83& 0.37 & 1&--&--&-- &-- &--&18$^5$&0.00&+0.76 \\

\hline

6402 (M14) &Int&-1.44$\pm$0.17 &-1.32&-1.17&0.53$\pm$0.07 &24&-1.23$\pm$0.21&-1.12&-1.12&0.58$\pm$0.05&36
&32&0.57&+0.65 \\

6779 (M56) &Int&-1.97$^3$&-2.05&-1.74&0.53 &1 &-1.96&-2.03&-2.05&0.51&1
&34&0.26&+0.98\\
\hline
6388  &III&-1.35$\pm$0.05&-1.23&-1.00&0.53$\pm$0.04&2&-0.67$\pm$0.24&-0.64&-0.56 &0.61$\pm$0.07&6
&12&0.40&--1.00 \\
6441  &III&-1.35$\pm$0.17&-1.23&-0.80&0.43$\pm$0.08&7&-1.02$\pm$0.34&-0.82&-1.00 &0.55$\pm$0.08&8
&13 &0.51&--0.73\\

\hline

\end{tabular}
\end{turn}

\center{\quad Notes: $^1$ Quoted uncertainties are 1-$\sigma$
errors calculated from the scatter in the data for each cluster. The number of stars
considered in the calculations is given by N. $^2$. The
only RRL V1 is probably not a cluster member. $^3$ This value has a -0.21 dex added, see
$\S$ \ref{sec:FeHMv} for a discussion. $^4$. Our calculation.  $^5$ Based on
one light curve not fully covered. $^6$ Metallicity and $M_V$ taken from the compilation of \citet{Kains2012}.\\

\quad References are the source of the Fourier coefficients: 1. \citet{Arellano2013a}; 2. \citet{Kains2012}; 3.
\citet{Arellano2014}; 4.  \citet{Arellano2018a}; 5. \citet{Kains2015}, 6. \citet{Arellano2011}; 7. \citet{Arellano2010}; 8. \citet{Arellano2008a}; 9. \citet{Arellano2013b}; 10. \citet{Yepez2020}; 11. \citet{Arellano2008b}; 12. \citet{Pritzl2002}  ; 13. \citet{Pritzl2001}; 14. \citet{Bramich2011}; 15. \citet{Arellano2006}; 16. \citet{Lazaro2006}; 17. \citet{Kains2013} ; 18. \citet{Figuera2013}; 19. \citet{Arellano2016}; 20. \citet{Arellano2015b}; 21. \citet{Tsapras2017}; 22. \citet{Deras2018}; 23. \citet{Walker1998}; 24. \citet{Cacciari2005}; 25. \citet{Arellano2019}; 26. \citet{Yepez2018} ;27. \citet{Arellano2018b}; 28. \citet{Yepez2019}; 29. \citet{Deras2019}; 30. \citet{Deras2020}; 31. \citet{Arellano2020}; 32. \citet{Yepez2022}; 33. \citet{Rojas2021}; 34. \citet{Deras2022}
}

\end{center}
\end{table*}

\section{Calculation of $M_V$ and \FeH}
\label{sec:Four}

Our approach to the calculation of mean $M_V$ and [Fe/H] for each GC in the sample has been through the RRL light curve Fourier decomposition, and the application of {\it ad hoc} well tested semi empirical calibrations. The Fourier decomposition of the RRL light curves is performed by fitting the
observed light curve in $V$-band with a Fourier series model of the form:

\begin{equation}
\label{eq.Foufit}
m(t) = A_0 + \sum_{k=1}^{N}{A_k \cos\ ({2\pi \over P}~k~(t-E) + \phi_k) },
\end{equation}

\noindent
where $m(t)$ is the magnitude at time $t$, $P$ is the period, and $E$ is the epoch. A
linear
minimization routine is used to derive the best-fit values of the 
amplitudes $A_k$ and phases $\phi_k$ of the sinusoidal components. 
From the amplitudes and phases of the harmonics in eq.~\ref{eq.Foufit}, the 
Fourier parameters, defined as $\phi_{ij} = j\phi_{i} - i\phi_{j}$, and $R_{ij} =
A_{i}/A_{j}$, are computed. 

Subsequently, the low-order Fourier parameters can be used in combination with
semi-empirical calibrations to calculate [Fe/H] and $M_V$ for
each RRL and hence the mean values of the metallicity and absolute magnitude for the RRL population in the host
cluster.

The specific calibrations and zero points used for RRab and RRc stars for this purpose are described in $\S$ \ref{calibrations}. Numerous
Fourier decompositions of RRL light curves can be 
found in the literature. However, over the years, each author has used different
calibrations and zero points to estimate $M_V$ and [Fe/H]. Our group has also
used slightly different equations in the earlier papers but in
the work by \citet{Arellano2010} zero points of the $M_V$ calibrations (see their $\S$ 4.2), were discussed and adopted, and we have used them subsequently. In the present paper we have recalculated $M_V$ and [Fe/H] for all clusters in the sample, using the calibrations described in the following section.

The final values found for $M_V$ and [Fe/H], the later expressed in the three different scales defined in $\S$ \ref{calibrations}, are listed in Table \ref{MV_FEH:tab}, which
is organized by Oosterhoff types; Oosterhoff (1939, 1944) realized that the periods of
fundamental-mode RRL, or RRab stars in a given cluster, group around two values;
0.55d (Oosteroff type I or OoI) and 0.65d (Oosteroff type II or OoII). OoI clusters
are systematically more metal-rich than OoII clusters. A third Oosterhoff class 
(OoIII) (Pritzl et al. 2000), which presently contains only two GCs, NGC 6388 and NGC
6441, is represented by very metal-rich systems where the periods of their RRab
stars average about 0.75d. A few clusters have been classified as of the intermediate type or OoInt, since the average periods of their RRab stars and their mean [Fe/H] fall between those of type I and type II clusters; it has been argued that OoInt clusters may be associated to an extragalactic origin \citep{Catelan2009} due to their similarity to dSph galaxies, satellites of the Milky Way, and their respective clusters.  In Table \ref{MV_FEH:tab} we include
of 16 OoI, 14 OoII, 2 OoIII and 2 OoInt clusters. The calculations have been
performed independently for RRab and RRc stars. For clusters with differential
reddening, i.e. NGC 1904, NGC 3201, NGC 6333 and NGC 6401, care has been taken
in calculating the individual reddening for each RRL. The interested reader is
referred to the original papers for a detailed discussion on that subject. 

\subsection{[Fe/H] and $M_V$ calibrations}
\label{calibrations}
For the calculation of [Fe/H] we adopted the following calibrations:

\begin{eqnarray}
\label{eq:RR0_Fe}
{\rm [Fe/H]}_{\rm J}= - 5.038 - 5.394 P + 1.345 \phi^{(s)}_{31},
\end{eqnarray}

\begin{eqnarray}
\label{eq:RR1_Fe}
{\rm [Fe/H]}_{\rm ZW} = 52.466 P^2 - 30.075 P + 0.131 \phi^{(c)2}_{31}  \nonumber \\
 - 0.982 \phi^{(c)}_{31} - 4.198 \phi^{(c)}_{31} P + 2.424
\end{eqnarray}

\noindent
from Jurcsik \& Kov\'acs (1996) and Morgan et al. (2007) for RRab and RRc stars,
respectively. In the above equations, $\phi^{(c)}$ and $\phi^{(s)}$ are the phases calculated either on a cosine or a sine series respectively, and they are correlated as $\phi^{(s)} = \phi^{(c)} - \pi$. The iron abundance on the Jurcsik \& Kov\'acs (1996) scale can be
converted into the Zinn \& West (1984) scale using the equation
[Fe/H]$_{\rm J}$ = 1.431[Fe/H]$_{\rm ZW}$ + 0.88 (Jurcsik 1995). Then, the [Fe/H]$_{\rm ZW}$ was transformed into the spectroscopic scale [Fe/H]$_{\rm UV}$ defined by \citet{Carretta2009} from high resolution spectroscopic determinations of the iron abundance, via the relation: [Fe/H]$_{\rm UV}$= $-0.413$ + 0.130~[Fe/H]$_{\rm ZW} - 0.356$~[Fe/H]$_{\rm ZW}^2$.

\citet{Nemec2013} calculated non-linear calibrations of [Fe/H] in terms of $\phi_{31}$ and pulsating period, using as calibrators the iron to hydrogen abundances of 26 RRab and 110 RRc stars calculated from high dispersion spectroscopy. For the RRc stars he added four RRc stars to the original 106 used by \citet{Morgan2007} (eq. \ref{eq:RR1_Fe}), and removed {\it a posteriori} nine outlier stars. Nemec's calibrations for the RRab and RRc stars are respectively of the form;

\begin{eqnarray}
{\rm[Fe/H]_{N}}=-8.65-40.12P+5.96\phi_{31}^{(s)}(K) \nonumber \\
+6.27\phi_{31}^{(s)}(K)P-0.72\phi_{31}^{(s)}(K){^2},
\label{NemAB}
\end{eqnarray}

\noindent
where $\phi_{31}^{(s)}(K)$=$\phi_{31}^{(s)}+0.151$ is given in the $Kepler$ scale \citep{Nemec2013}, and

\begin{eqnarray}
{\rm [Fe/H]_{N}}=1.70-15.67P+0.20\phi_{31}^{(c)}-2.41\phi_{31}^{(c)}P \nonumber \\
+18.0P^2  +0.17\phi_{31}^{(c)^2}.
\label{NemC}
\end{eqnarray}.

As pointed out by \citet{Nemec2013}, since the above calibrations are based on high resolution spectroscopic determinations of [Fe/H], the derived values [Fe/H]$_{\rm N}$ are on the UV scale of \citet{Carretta2009}. Thus, they should be comparable to the values [Fe/H]$_{\rm UV}$, a point on which we shall comment below.

For the calculation of $M_V$ we adopted the calibrations:

\begin{equation}
\label{eq.RR0_Mv}
M_V= -1.876~log P -1.158 A_1 + 0.821 A_3 + 0.41,
\end{equation}

\begin{equation}
\label{eq.RR1_Mv}
M_V= -0.961 P - 0.044 \phi^{(s)}_{21} -4.447 A_4 + 1.061,
\end{equation}

\noindent
from \citet{Kovacs2001}  and \citet{Kovacs1998} for the RRab and RRc stars,
respectively. The zero points of eqs. \ref {eq.RR0_Mv} and \ref {eq.RR1_Mv} have been
calculated to scale the
luminosities of RRab and RRc stars to the distance modulus of 18.5 mag for the Large
Magellanic Cloud (LMC) (see the discussion in $\S$ 4.2 of \citet{Arellano2010}.

In Table \ref{MV_FEH:tab} we list the globular clusters studied by our team and the resulting [Fe/H] in the scales of \citet{Zinn1984}, \citet{Carretta2009} and \citet{Nemec2013}, i.e. 
[Fe/H]$_{\rm ZW}$, [Fe/H]$_{\rm UV}$ and [Fe/H]$_{\rm N}$ and
$M_V$, estimated via the Fourier decomposition of the light curves of the RRab and RRc
stars. To this end, we have taken the Fourier parameters published in the original papers and applied the above calibrations for the sake of homogeneity.
We have also included the two metal-rich clusters NGC 6388 and NGC 6441 studied by
\citet{Pritzl2002,Pritzl2001}, NGC 1851 \citep{Walker1998}  and NGC 5272 (M3) \citep{Cacciari2005}  since the light curve Fourier decomposition parameters are available in those papers. In order to increase the sample, the clusters NGC 5286, NGC 6266 and NGC 6809, have been added. For these,   Fourier-based physical parameters have been reported by \citet{Zorotovic2010}, \citet{Contreras2010} and \citet{Olech1999} respectively and the results have been duly transformed to the proper scales by \citet{Kains2012}. In all these calculation of the Fourier-based physical parameters, we have systematically avoided clear Blazhko variables or any amplitude modulated stars. All the relevant papers are recorded in the notes to Table \ref{MV_FEH:tab}.

The use of the
above equations and their zero points form the basis of the discussion of the
$M_V$-[Fe/H]
relation and the cluster distances on a homogeneous scale, which we present in the following sections.

\section{Vindication of the photometric approach to the metallicities}

\label{VINDI}

There is absolutely no doubt that the most precise approach to the determination of metallicities of heavenly bodies is via high-resolution spectroscopy. The practical limitations to that technique are several however; to reach deep in magnitude, typical of the HB in most globular clusters, long exposure times with large telescopes are required making it unaffordable. The spectroscopic values for 19 clusters listed by \citet{Carretta2009} were obtained in numeorus elaborated previous papers cited by these authors. The analyses were carried on luminous red giants of $V \sim 14-16$ mag, i.e. 2-3 magnitudes brighter than RRL stars at the HB, and after an enormous compromise of observational and computational resources. While this situation may change with the advent of in-orbit high resolution spectrographs, the competition for access to the instrumentation will likely remain tough.
The photometric approach to the metallicity and luminosity calculation in RRL stars was envisaged in the 1980's \citep{SiminTeays1982}, and became a popular alternative since it reaches as low $V \sim 20$ mag with sufficient accuracy with very short exposure times with 1-2m-class telescopes, enabling the access to larger samples of clusters. The Fourier decomposition approach was further developed to produce the calibrations exposed in $\S$ \ref{sec:Four} en employed in this paper.

\begin{figure*}
\begin{center}
\includegraphics[width=15.0cm, height=15.0cm]{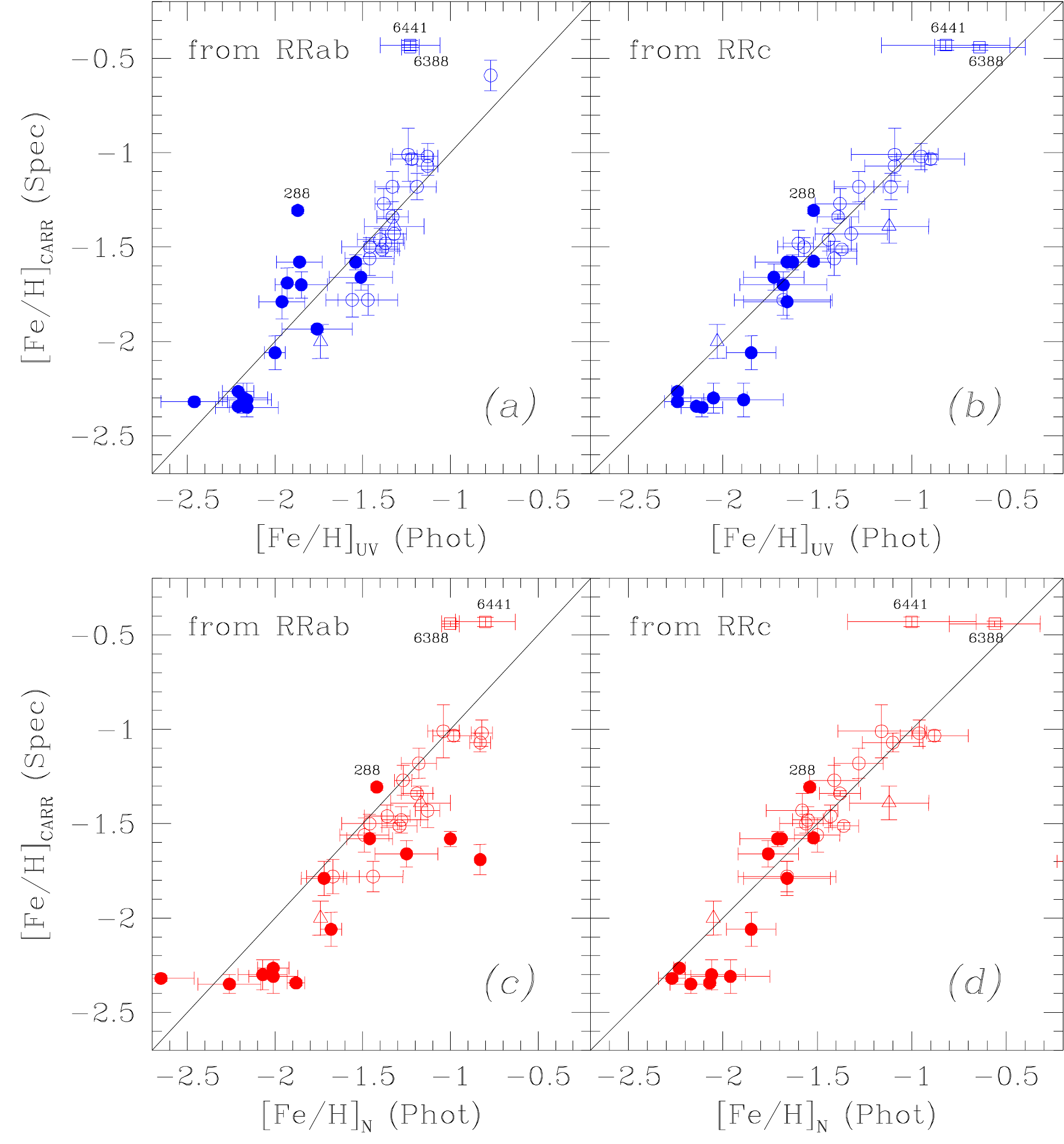}
\caption{Comparison of the Fourier based [Fe/H] (Table \ref{MV_FEH:tab}), with those from  high-resolution spectroscopy based values from \citet{Carretta2009}. Filled and empty circles represent OoII and OoI clusters respectively. Open triangles and squares represent OoInt and OoIII clusters. See $\S$ \ref{VINDI} for a discussion.}
\label{PhoVSspec}
\end{center}
\end{figure*}

Our goal in this section is to compare the photometric values reported in this work with the spectroscopic values of \citet{Carretta2009}. Fig \ref{PhoVSspec} shows the photometric based values of [Fe/H] in the UV and Nemec scales (Table \ref{MV_FEH:tab}) for the RRab and RRc stars, plotted versus the spectroscopic values given by \citet{Carretta2009}. In panels $(a)$ and $(b)$ it is clear that the comparison is satisfactory for the case of $\rm [Fe/H]_{UV}$. However, for $\rm [Fe/H]_{N}$ the RRab calibration seems to systematically overestimates the metallicity relative to the spectroscopic values. The iron values from Nemec's calibration for RRc stars also compare well with the spectroscopic values. There is a mild suggestion in both panels $(b)$ and $(d)$ that the calibrations for the RRc stars of equations \ref{eq:RR1_Fe} and \ref{NemC}, which are in fact based on the same set of calibrators, may require a small adjustment of about -0.2 dex for iron values  smaller than -2.0.

\section{The $M_V$-[{F\MakeLowercase{e}/H}] ~correlation}

It has been argued that eq. \ref{eq:RR0_Fe} overestimates [Fe/H] for metal-poor clusters.
This problem has been addressed by  \citet{Jurcsik1996}, \citet{Schwarzenberg1998}, \citet{Kovacs2002}, \citet{Nemec2004} and \citet{Arellano2010}.
It is difficult to quantify a correction to be applied, and it is likely also
a function of the metallicity, however, empirical estimations in the above papers
point to a value between --0.2 and  --0.3 dex on the scale of eq. \ref{eq:RR0_Fe}. We
have adopted --0.3 dex, which on the ZW scale is equivalent to --0.21 dex. Equally difficult it is to define a value of [Fe/H]$_{\rm ZW}$ bellow which the corrections should be applied. Guided by the metallicity values of globular clusters in the spectroscopic scale of \citet{Carretta2009}, we estimated that a reasonable limit would be [Fe/H]$_{\rm ZW} < -1.7$
Therefore,
the values listed in Table \ref{MV_FEH:tab} for clusters with [Fe/H]$_{\rm ZW} <$ --1.7
dex
were obtained by adding --0.21 dex to the value of [Fe/H]$_{\rm ZW}$ found via eq.
\ref{eq:RR0_Fe}. As a consequence the [Fe/H]$_{\rm UV}$ values for these clusters is also affected by this correction. Note that the good comparison between the photometric [Fe/H]$_{\rm UV}$ and the spectroscopic values displayed in Fig \ref{PhoVSspec} $(a)$ was obtained after the application of the correction above. 

\begin{figure*}
\begin{center}
\includegraphics[width=16.0cm, height=9.0cm]{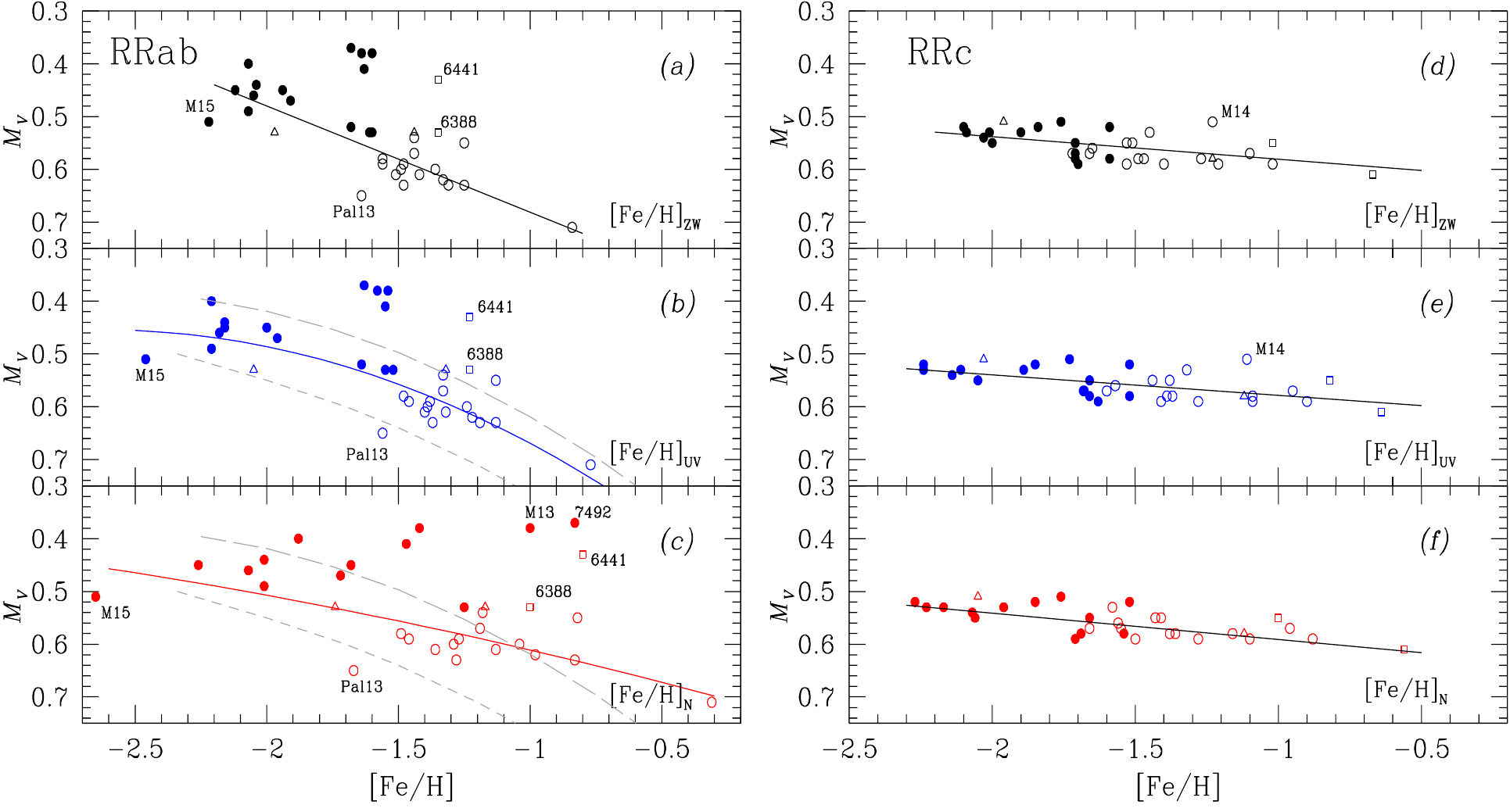}
\caption{The [Fe/H] versus $M_V$ correlations for RRab and RRc stars. The involved metallicity scale is, from top to bottom panels, [Fe/H]$_{\rm ZW}$, [Fe/H]$_{\rm UV}$ and [Fe/H]$_{\rm N}$. Filled and open circles represent OoII and OoI clusters respectively. Open triangles and squares represent OoInt and OoIII clusters. All fits have been weighted by the number of stars included in each cluster. In the left panel, the gray curves are the theoretical predictions of \citet{Cassisi1999} (long dash) and \citet{Vandenberg2000} (short dash), which are notoriously similar to the photometric solution in panel {\it (b)}.}
\label{FeHMv}
\end{center}
\end{figure*}

In Fig. \ref{FeHMv} we show the distribution of clusters in the $M_V$-[Fe/H]
plane obtained from the RRab stars
(left panel) and the RRc stars (right panel). In each of these panels we display the resulting distributions for the three involved scales [Fe/H]$_{\rm ZW}$, [Fe/H]$_{\rm UV}$ and [Fe/H]$_{\rm N}$. In the middle and bottom boxes, for the spectroscopic scales [Fe/H]$_{\rm UV}$ and [Fe/H]$_{\rm N}$, we include as
reference, in gray colour, two theoretical, non-linear, versions of the $M_V$-[Fe/H] relation of
Cassisi 
et al. (1999) and VandenBerg et al. (2000). We shall discuss these correlations separately for the RRab and RRc stars.

\subsection{From RRab stars in globular clusters}

The trend between [Fe/H] in all scales and $M_V$ is evident, as much as it is the large scatter. There are a few outliers, labeled in the figure, that were not considered in the calculations of the fitted regressions (with the exception of M15). However, some evidence of non-linearity is suggested in the central and bottom panels of Fig. \ref{FeHMv}, particularly oriented by the presence of M15 that is the most metal-poor cluster in the sample, hence its relevance. It is also worth noting that \citet{Nemec2013} calibration, eq. \ref{NemAB}, includes a wider selection of calibrators  with metallicities below -2.0, and as low as [Fe/H] $\sim -2.68$ (for star X Ari). Since the number of stars involved in the calculation of the physical parameters varies from cluster to cluster, all the fits below have been weighted by $1/(\sigma_i^2/N_i) $. The quadratic fits for [Fe/H]$_{\rm UV}$ and [Fe/H]$_{\rm N}$ are of the form:

\begin{eqnarray}
\label{FEHuv}
M_V= 1.016(\pm 0.170) + 0.428(\pm 0.207) \rm[Fe/H]_{\rm UV} + \nonumber \\
+0.081(\pm 0.060) \rm[Fe/H]_{\rm UV}^2,
\end{eqnarray}
\noindent
with an rms=0.060 mag, and

\begin{eqnarray}
\label{FEHN}
M_V= 0.740(\pm 0.056) + 0.141(\pm 0.077) \rm[Fe/H]_{\rm N} + \nonumber \\
+ 0.013(\pm 0.026) \rm[Fe/H]_{\rm N}^2,
\end{eqnarray}
\noindent
with an rms=0.060 mag.

The quadratic empirical solutions for [Fe/H]$_{\rm UV}$, shown in Fig. \ref{FeHMv} $(b)$, is  remarkably similar to the theoretical predictions of \citet{Vandenberg2000} and \citet{Cassisi1999}, in shape and luminosity level. To our knowledge, this is the first empirical solution that reproduce the theoretical predictions of the non-linear nature of the correlation, which it has likely been enabled by the homogeneous treatment of a large number of clusters, and the distinction of RRab and RRc stars.

We call attention towards the inclusion of the metal rich cluster NGC 6366 (--0.77,0.71) in the UV correlation for the RRab, in spite of its metallicity been derived from a single star that might not be a cluster member \citep{Arellano2008a}. However excluding it or employing the value [Fe/H]=--0.59 listed by \citet{Harris1996} makes no significant variation in the correlation.

\subsection{From RRc stars in globular clusters}

The mean [Fe/H] and $M_V$ determined from the RRc stars in the family of studied globular clusters are
correlated as shown in the right panel of Fig. \ref{FeHMv}. Immediate differences are seen when compared to the cases from the RRab stars: the slopes are milder, the distributions in the three metallicity scales are all very similar, the correlations are strikingly tight, in spite of which no suggestion of a non-linear correlation is evident. It should also be noted that the Oo-int (triangles) and the OoIII clusters (squares) follow the trends well.

The linear correlations for the 
[Fe/H]$_{\rm UV}$ and [Fe/H]$_{\rm N}$ can be expressed as:

\begin{eqnarray}
\label{FEHUVc}
M_V= 0.034 (\pm 0.009)\rm[Fe/H]_{\rm UV} + 0.601 (\pm0.015),
\end{eqnarray}

\noindent
and

\begin{eqnarray}
\label{FEHNc}
M_V= 0.050 (\pm 0.004) \rm[Fe/H]_{\rm N} + 0.641 (\pm0.006), 
\end{eqnarray}

\noindent
with an rms=0.022 mag.

Eqs. \ref{FEHUVc} and \ref{FEHNc} are basically identical. The reason is that, although the values of [Fe/H]$_{\rm UV}$ and [Fe/H]$_{\rm N}$, come from different formulations (eqs. \ref{eq:RR1_Fe} and \ref{NemC}), both calibrations come essentially from the same set of calibrator stars, since \citet{Nemec2013} took the calibrators from \citet{Morgan2007}, and added four stars, for a total sample of 101 stars.

The remarkable difference of the cluster distribution on the $M_V$-[Fe/H] plane for the luminosity and metallicity determinations from the Fourier decomposition for RRab and RRc stars, does require some considerations. Naturally one may wonder if this is an artifact of the calibrations employed to transform Fourier parameters into physical parameters. However, the good agreement of the photometric metallicities [Fe/H]$_{\rm UV}$ and the spectroscopic values (Fig. \ref{PhoVSspec} $(a)$), and also the good cluster distance agreement with independent high-quality determinations presented below in $\S$ \ref{sec:distances}, offer support to the photometric calibrations given in $\S$ \ref{sec:Four} and their results in Table \ref{MV_FEH:tab}. In our opinion, the run of [Fe/H] with $M_V$ and the scatter seen in RRab stars are a consequence of the interconnection of the following: RRab stars are larger amplitude variables with a more complex light curve morphology, often the light maximum is very acute, and they are prone to display amplitude and phase modulations, therefore their Fourier decomposition is subject to further uncertainties as they require a larger number of harmonics for a proper representation. Their larger periods may also limit a proper coverage of their pulsating cycle. Also, RRab stars in a given cluster may display small evolutionary stage differences, spreading a luminosity range. These circumstances have their impact in the calibration of the Fourier parameters and in the resultant scatter in the $M_V$-[Fe/H] plane. On the other hand, RRc stars have simpler light curves, mostly sinusoidal, and are more concentrated towards the ZAHB, thus their Fourier and physical parameters tend to be better correlated.

In summary, the HB luminosity-metallicity correlation seen from the RRab stars is steeper (a fact that had already been reported by ABG17), more scattered and non-linear, whereas from the RRc stars the relation is milder but better defined and linear. RRab and RRc stars should not be mixed for the purpose of studying or applying the correlation as a distance indicator instrument. Therefore, for the sake of estimating a globular cluster distance from its RRL, given its metallicity, one should prefer the RRc stars whenever possible and either eqs. \ref{FEHUVc} or \ref{FEHNc}. 

\subsection{The role of the HB structure parameter}

\citet{Demarque2000}, have argued on theoretical grounds that the overall structure of the HB plays a relevant role and may be interconnected with the HB luminosity and the metallicity of the parental globular cluster. Here we explore the role of the HB type parameter, or Zinn-Lee parameter, defined as $\mathcal L$ = $(B-R) / (B+V+R)$ \citep{Zinn1986,Lee1990}, from empirical arguments. $B, V$ and $R$ represent the number of stars to the blue of the instability strip (IS), the number of RRL stars, and to the red of the IS respectively. For a better calculation of $\mathcal L$, it is convenient to include, as far as possible, only cluster member stars in the counting.
Since the list of $\mathcal L$ values for a large number of clusters presented by \citet{Torelli2019} are the result of membership considerations, we have adopted them for the subsequent analysis. When a cluster is not included in the above list we used the value reported by \citet{Catelan2009}. The one exception is NGC 7079 (M2). For this cluster, the reported value is $\mathcal L$=+0.96. The analysis of the projected positions and proper motions available in the $Gaia$-eDR3 data base, and the application of  the method of \citet{Bustos2019}, kindly performed by Dr. Bustos Fierro, renders a CMD of very likely cluster member stars, with a substantial population of red HB stars, for $\mathcal L$=+0.34 which shall be adopted. For a few other clusters where similar analysis have been carried out, $\mathcal L$ values close to those  of \citet{Torelli2019} were found.

\begin{figure}
\begin{center}
\includegraphics[width=8.0cm, height=8.0cm]{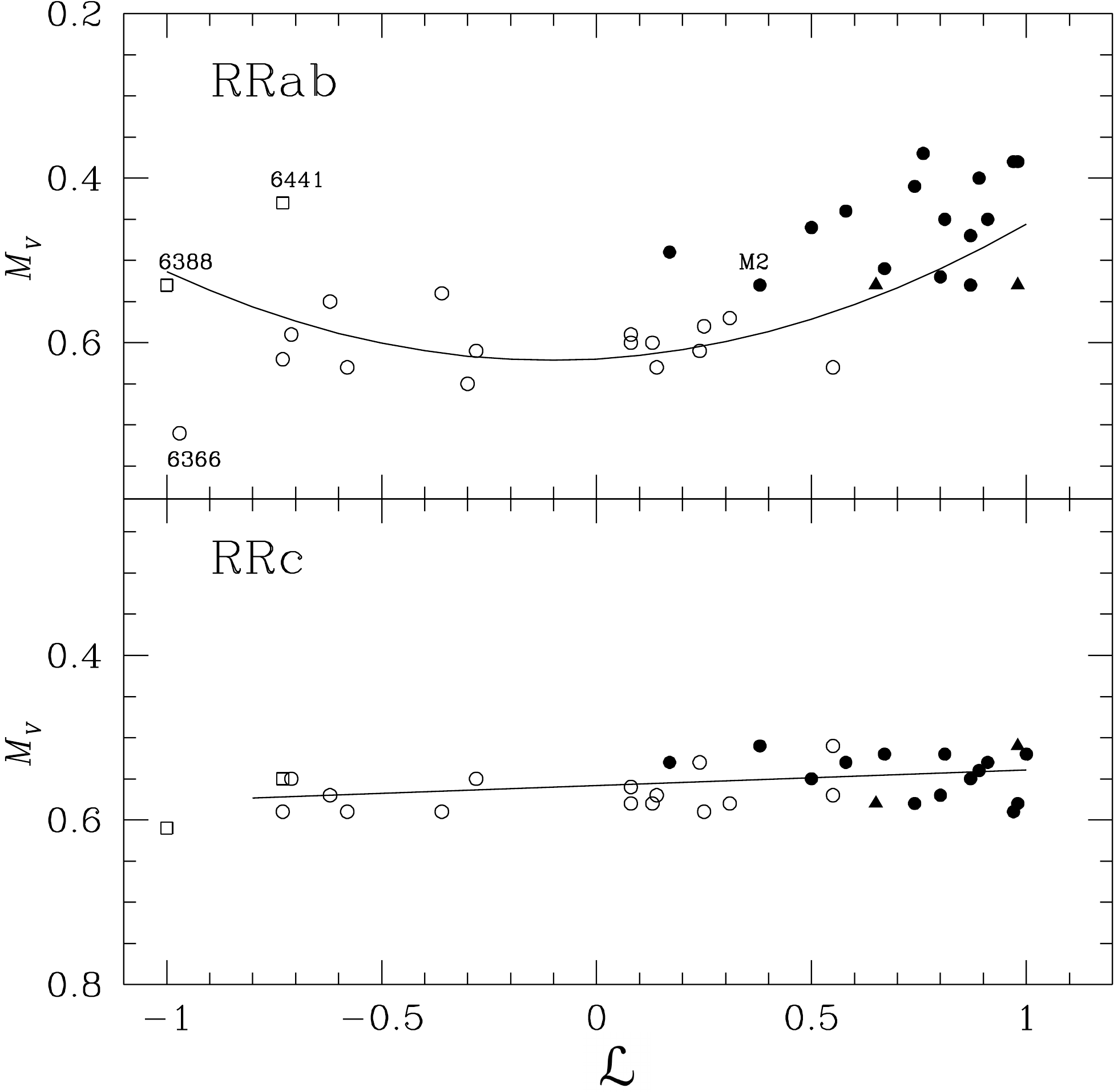}
\caption{The HB structure parameter $\mathcal L$ vs. mean $M_V$ for a family of clusters coded as in Fig. \ref{PhoVSspec}. Two open squares for the OoIII clusters were not considered in the weighted fits.}
\label{LvsMv}
\end{center}
\end{figure}

Fig. \ref{LvsMv} illustrates the correlation between $\mathcal L$ and $M_V$, the later as obtained from the Fourier decomposition for RRab stars (top panel) and RRc stars (bottom panel). In the case of the RRab stars the correlation clearly shows a quadratic trend, once the two Oo-Int clusters, NGC 6388 and NGC 6441, were excluded. In the case of the RRc stars the correlation appears linear with a very mild slope. The corresponding fits in Fig. \ref{LvsMv} are of the form:

\begin{eqnarray}
\label{MvLab}
M_V= 0.620(\pm 0.011) - 0.029(\pm 0.018) {\mathcal L} \nonumber \\
- 0.135 (\pm0.036) {\mathcal L}^2,
\end{eqnarray}
\noindent
with and rms = 0.058, and

\begin{eqnarray}
\label{MvLc}
M_V= 0.558(\pm 0.006) - 0.019(\pm 0.013) {\mathcal L},
\end{eqnarray}
\noindent
with and rms = 0.026.

Considering the trends in Fig. \ref{FeHMv} for the RRab and RRc, both for the $\rm [Fe/H]_{UV}$ values (middle panels), and in Fig. \ref{LvsMv}, the dependence of $M_V$ on the metallicity [Fe/H] and HB structure parameter $\mathcal L$, can be expressed, for the RRab and RRc respectively as:

\begin{eqnarray}
\label{MVFELab}
M_V= \rm A + \rm B \rm[Fe/H]_{\rm UV} + C \rm[Fe/H]_{\rm UV}^2 + D \mathcal L \nonumber \\
+ E {\mathcal L}^2,
\end{eqnarray}

\noindent
with A=+1.096($\pm 0.141$), B=+0.519($\pm 0.172$), C=+0.119($\pm 0.050$), D=+0.006($\pm 0.014$), E=--0.111($\pm 0.029$), and rms = 0.053 mag.

\begin{eqnarray}
\label{MVFELc}
M_V= +0.609 (\pm 0.016) + 0.032 (\pm 0.009) \rm[Fe/H]_{\rm UV} \nonumber \\
+ 0.015 (\pm 0.011)\mathcal L,
\end{eqnarray}

\noindent 
with  rms = 0.024 mag.

The equivalent calibrations in terms of the metallicity in the scale of \citet{Nemec2013}, [Fe/H]$_{\rm N}$ are:

\begin{eqnarray}
\label{MVFELnab}
M_V= \rm A + \rm B \rm[Fe/H]_{\rm N} + C \rm[Fe/H]_{\rm N}^2 + D \mathcal L + E {\mathcal L}^2,
\end{eqnarray}

\noindent
with A=+0.720($\pm 0.082$), B=+0.130($\pm 0.098$), C=+0.033($\pm 0.029$), D=--0.043($\pm 0.020$), E=--0.145($\pm 0.030$), and rms = 0.055 mag.

\begin{eqnarray}
\label{MVFELnc}
M_V= +0.655 (\pm 0.019) + 0.063 (\pm 0.013) \rm[Fe/H]_{\rm N} \nonumber \\
+ 0.012 (\pm 0.009)\mathcal L,
\end{eqnarray}

\noindent 
with  rms = 0.019 mag.

For the calibration of eqs. \ref{MVFELab} and \ref{MVFELnab} from the RRab solutions, the terms involving $\mathcal L$ are small but significant. On the contrary, eqs. \ref{MVFELc} and \ref{MVFELnc} from the RRc solutions, the last term is insignificant, and in fact, for instance eqs. \ref{FEHUVc} and \ref{MVFELc} in the UV scale, or eqs. \ref{FEHNc} and \ref{MVFELnc} in the Nemec's scale, are, within the uncertainties, indistinguishable, confirming the linearity and sufficiency of an $M_V$-[Fe/H] relation for the RRc stars.

Therefore, the empirical $M_V$-[Fe/H] relation as worked out from the Fourier decomposition of RRab stars light curves, has turned out to be much more complex, with the metallicity and HB structure playing a measurable role, and eqs. \ref{MVFELab} and \ref{MVFELnab} are  a good representation. 

For the RRc stars, with simpler light curves and  generally more confined near the ZAHB, the $M_V$-[Fe/H] relation remains linear and simple, and the structure of the HB do not seem to play any pertinent role; eq. \ref{FEHUVc} and \ref {FEHNc}, or for any purpose eqs. \ref{MVFELc} and \ref{MVFELnc}, are good empirical calibrations, with a well established slope around 0.06.

\begin{figure}
\begin{center}
\includegraphics[width=8.0cm, height=12.5cm]{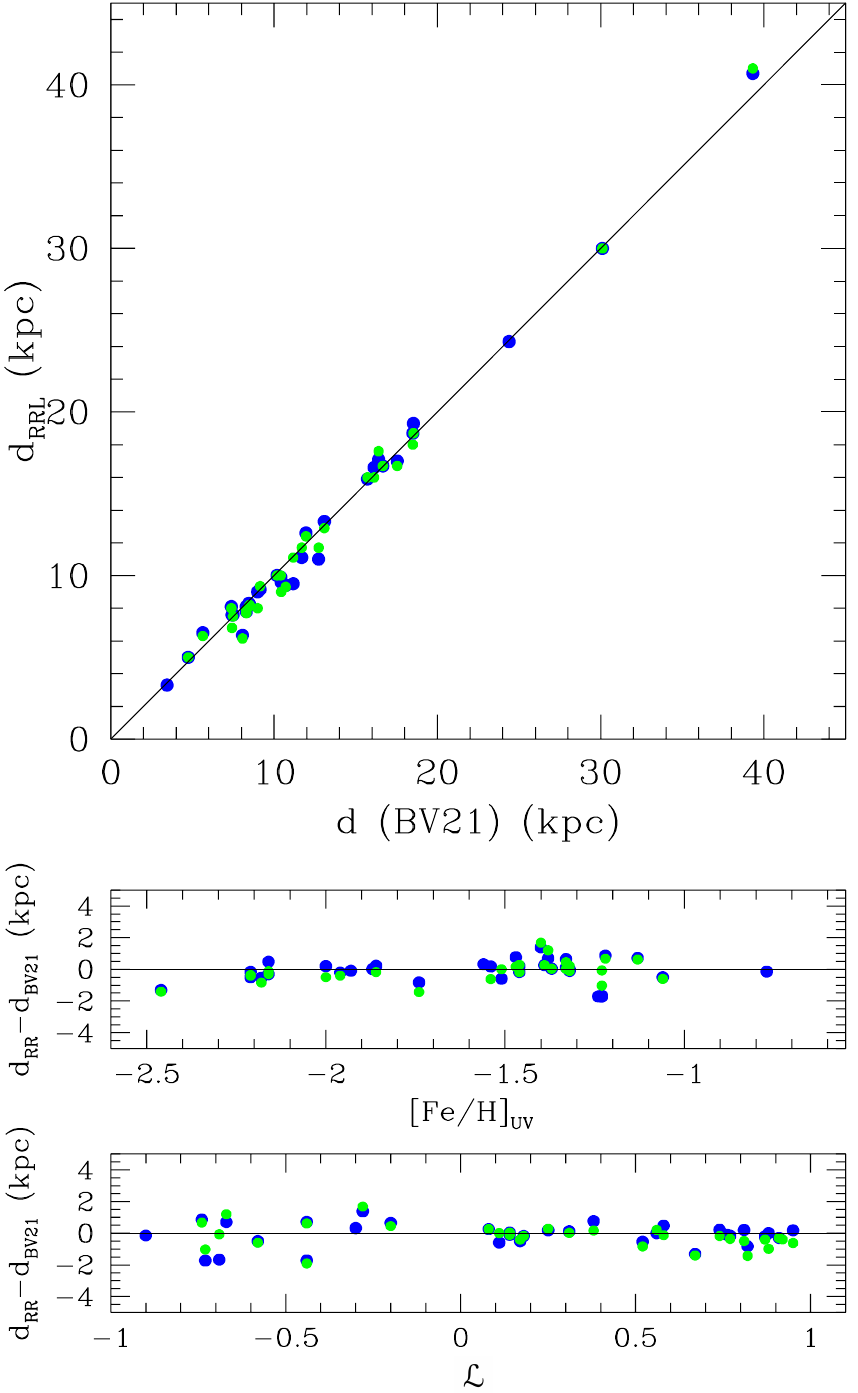}
\caption{Comparison of distances obtained from the RRL Fourier decomposition and those of BV21. Blue and green symbols are for distances derived from RRab and RRc Fourier light curve treatment, respectively.}
\label{DISTANCIAS}
\end{center}
\end{figure}

\section{Globular cluster distances}
\label{sec:distances}

Once the mean absolute magnitude $M_V$ for the HB is obtained for a given cluster, its distance can be estimated for an assumed value of $E(B-V)$. Using the values listed in Table \ref{MV_FEH:tab} the distances were calculated and are listed in Table \ref{T2:distance} as they were obtained either for the RRab or RRc stars from eqs. \ref{eq.RR0_Mv} and \ref{eq.RR1_Mv} respectively. We perform a comparison with accurate mean distances recently estimated by \citet{Baumgardt2021} (hereinafter BV21), calculated for a large sample of globular clusters using the data from $Gaia$-eDR3, $HST$ and selected literature distances.

It should be obvious from Table \ref{T2:distance} that the distances derived from our Fourier approach (columns 2 and 3) agree with those of BV21. Fig. \ref{DISTANCIAS} shows a graphical comparison and how the distance differences do not correlate neither with the metallicity nor the HB structure parameter. The Fourier and the BV21 distances differences are all smaller than 1.7 kpc and display a standard deviation of 0.7 kpc.

\begin{table*}
\scriptsize
\begin{center}

\caption{Distances for a sample of Globular Clusters estimated homogeneously
from the RRL stars light curve Fourier decompositions.}
\label{T2:distance}

\begin{tabular}{lccccccc}
\hline
GC &$d (kpc)$&$d (kpc)$& $d$ (kpc) &No. of & $d$ (kpc)&$E(B-V)$ &d (kpc)\\
NGC(M)  & (RRab)& (RRc) & (SX Phe) &SX Phe&  (SX Phe)& & \\
   &  &   &P-L AF11 &&P-L CS12 &&BV21\\
\hline
 288  &9.0$\pm$0.2 &8.0 &8.8$\pm$0.4&6&9.4$\pm$0.6&0.03 & 8.988\\
1261 &17.1$\pm$0.4&17.6$\pm$0.7&&&&0.01&16.400\\
1851 &12.6$\pm$0.2 &12.4$\pm$0.2&--&--& --&0.02&11.951\\
1904 (M79) &13.3$\pm$0.4&12.9&--& --&--&0.01& 13.078\\ 
3201 &5.0$\pm$0.2&5.0$\pm$0.1&4.9$\pm$0.3&16&5.2$\pm$0.4&dif& 4.737 \\ 
4147 &19.3 &18.7$\pm$0.5 &--& --&--&0.02&18.535\\   
4590 (M68)&9.9$\pm$0.3&10.0$\pm$0.2&9.8$\pm$0.5&6 &--&0.05& 10.404\\
5024 (M53)&18.7$\pm$0.4&18.0$\pm$0.5&18.7$\pm$0.6&13&20.0$\pm$0.8&0.02&18.498 \\ 
5053 &17.0$\pm$0.4 &16.7$\pm$0.4&17.1$\pm$1.1&12&17.7$\pm$1.2&0.02&17.537 \\
5272 &10.0$\pm$0.2&10.0$\pm$0.4&&&&0.01&10.175\\
5466 &16.6$\pm$0.2 &16.0$\pm$0.6&15.4$\pm$1.3&5&16.4$\pm$1.3&0.00&16.120 \\
5904 (M5) &7.6$\pm$0.2 &7.5$\pm$0.3&6.7$\pm$0.5&3&7.5$\pm$0.2&0.03&7.479 \\
6205 (M13) &7.6 &6.8$\pm$0.3&7.2$\pm$0.7&4&--&0.02&7.419 \\
6171 &6.5$\pm$0.3&6.3$\pm$0.2&&&&0.33&5.631\\
6229 &30.0$\pm$1.5  &30.0$\pm$1.1 &27.9 & 1&28.9&0.01&30.106 \\
6254 (M10) &--&4.7&5.2$\pm$0.3&15& 5.6$\pm$0.3&0.25&5.067\\
6333 (M9) &8.1$\pm$0.2 &7.9$\pm$0.3&--&--& --&dif&8.300\\
6341 (M92)&8.2$\pm$0.2 &8.2$\pm$0.4&--&--& --&0.02&8.501\\
6362 &7.8$\pm$0.1&7.7$\pm$0.2&7.1$\pm$0.2&6&7.6$\pm$0.2&0.09&8.300\\
6366 &3.3 &--&--&--& --&0.80&3.444\\
6388 &9.5$\pm$1.2&11.1$\pm$1.1 &--&--&--&0.40&11.171\\ 
6401 &6.35$\pm$0.7 &6.15$\pm$1.4 &--&--&--&dif& 8.064\\
6402 (M14)&9.1$\pm$0.9 &9.3$\pm$0.5&--&--&--&0.57& 9.144\\
6441 &11.0$\pm$1.8 &11.7$\pm$1.0 &--&--& --&0.51&12.728\\ 
6712 &8.1$\pm$0.2&8.0$\pm$0.3&&&&0.35&7.382\\
6779 (M56)&9.6&9.0& &--&--&0.26&10.430\\
6934 &15.9$\pm$0.4&16.0$\pm$0.6&15.8& 1&18.0&0.10&15.716 \\
6981 (M72)&16.7$\pm$0.4&16.7$\pm$0.4&16.8$\pm$1.6&3&18.0$\pm$1.0&0.06&16.661 \\
7006 &40.7$\pm$1.6&41.0$\pm$1.6&&&&0.08&39.318\\
7078 (M15)&9.4$\pm$0.4&9.3$\pm$0.6&--& --&--&0.08& 10.709\\
7089 (M2) &11.1$\pm$0.6&11.7$\pm$0.02&--& --&--&0.06& 11.693\\
7099 (M30)&8.32$\pm$0.3 &8.1&8.0&1&8.3&0.03&8.458 \\
7492 &24.3&--&22.1$\pm$3.2&2& 24.1$\pm$3.7&0.00&24.390\\
Pal 13 &23.8$\pm$0.6&--&&&&0.10&23.475\\
\hline
\end{tabular}
\end{center}
\end{table*}

The agreement is remarkably good considering that the distance determinations come from completely independent approaches. We note that the distances obtained from the $M_V$ calibrations for RRab and RRc stars, are also independent, as they come from different and independent calibrations. This gives further support to the $M_V$ calibrations of eqs. \ref{eq.RR0_Mv} and \ref{eq.RR1_Mv} and to their zero points \citep{Arellano2010}.

The accurate distances of individual globular clusters obtained from the RRL stars listed in Table \ref{DISTANCIAS}, can serve as a frame of comparison of other independent methods to calculate cluster distances. Given that SX Phe stars are common in globular clusters, they can be used as secondary distance indicators through their well established P-L relation, of which, however, different calibrations are found in the literature. We shall explore the consistency of the results. We considered the calibrations of  \citet{Arellano2011} (AF11), and \citet{CohenSara2012} (CS12). The resulting distances from these two calibrations of 13 clusters with well observed SX Phe stars are listed in columns 4 and 6 of Table \ref{T2:distance}, column 5 indicates the number of member SX Phe stars available in each cluster. We should emphasize that the two calibrations lead to distances agreeing within 1.3 kpc, except for NGC 6934 where the distance differences is $\sim$ 2.0 kpc for the CS12 calibration. The SX Phe distances agree with the RRL results well within 1 kpc, i.e. the average RRL and SX Phe distance match in average by $\sim$ 4\% of the corresponding distance. 

\section{Variable Stars in our sample of Globular Clusters}
\label{VarsGC}

Once a CCD time-series photometry is performed on a given cluster, a by-product of the exercise is the discovery of previously undetected variables. In the work carried by our group, we have systematically searched for variables via an variety of approaches described in the individual papers, e.g. \citet{Arellano2013a}. In Table \ref{NoVARS} we summarize the number of variables, and their types, known in the globular clusters of our sample, noting the ones found by our work.
We have found 326 new variables in the field of the clusters, 23 of them are either considered field stars or have not been classified. The most numerous families are in order RRab, RRc, SR, SX Phe, eclipsing binaries, CW and double mode RRd stars. The total number of variables detected in these clusters is 2047 but only 1886 are likely truly cluster members. Thus, about 16\% of the variables in this sample of clusters has been found by our \emph{VI} CCD time-series imaging program.

\begin{table*}
\scriptsize
\begin{center}
\caption{Number of presently known variables per cluster for the most common variable
types, in a sample of globular clusters studied by our group$\dagger$.}
\label{NoVARS}
\begin{tabular}{lccccccccccc}
\hline
GC & RRab & RRc &RRd & SX Phe& Binaries& CW-(AC)-RV& SR, L,M& spotted&unclass& Total per cluster & Ref.\\
NGC (M) & & && & & & & &others $*$& &\\
\hline
288  &0/1&0/1&0/0&0/6 &0/1&0/0&0/1&&&0/10&1\\
1261 &0/16 & 0/6 & 0/0 & 0/3 & 0/1 & 0/0 & 0/3 &&& 0/29 & 22\\
1904 (M79)&0/6&1/5&0/0&0/5&0/1&0/1&0/14&&0/1&1/32&\\ 
3201 &0/72&0/7&0/0&3/24&0/11&0/0&0/8&0/2&0/7&3/124&3\\ 
4147 &0/5 &0/19 &0/1 &0/0 &0/14 &0/0 &2/2 &&0/3&2/41&4,23,35\\  
4590 (M68)&0/14 &0/16&0/12&4/6&0/0&0/0&0/0&&1/2&4/48&5\\
5024 (M53)&0/29&2/35&0/0&13/28&0/0&0/0&1/12&&&16/104&6,7\\ 
5053 &0/6 &0/4&0/0&0/5&0/0&0/0 &0/0&&&0/15&8\\
5466 &0/13 &0/8&0/0&0/9&0/3&0/1&0/0&&2/2&0/34&9\\
5904 (M5)&2/  89&1/40&0/0&1/6&1/3&0/2 &11/12 &&0/1&16/152&17,18\\
6171 (M107)&0/15 &0/6 &0/0 &0/1 &0/0 &0/0 &2/3 &&0/3&2/25 &24\\
6205 (M13)& 0/1 & 1/7& 2/2 & 2/6 & 1/3 & 0/3 &3/22 &&0/4 & 9/44 &25\\
6229 &10/42 &5/15 &0/0 &1/1&0/0 &2/5&6/6 &&0/1&24/69&19\\
6254 (M10)&0/0 & 0/1 &0/0 & 1/15 & 2/10 & 0/3 & 0/5 &&0/2 & 3/34 & 26\\
6333 (M9)&0/8 &2/10&1/1&0/0&3/4&1/1&5/6&&3/4&12/30&10\\
6341 (M92)&0/9 &0/5 & 1/1 & 1/6 & 0/0 & 0/1 & 1/1 &&0/6 & 3/23 & 27\\
6362 &0/16 &0/15 &1/3& 0/6 & 0/12 & 0/0&0/0 &0/3&0/22& 1/55 &28\\
6366 &0/1 &0/0&0/0&1/1 &1/1&1/1&3/4&&&6/8&11\\
6388&1/14&2/23&0/0&0/1&0/10 &1/11&42/58&&&46/117&21\\
6397 &0/0 &0/0&0/0&0/5 &0/15&0/0&0/1&&0/13&0/21&29\\
6401 &6/23&6/11&0/0&0/0&0/14&0/1&3/3&&14/14&15/52&20\\
6402 (M14)&0/55 &3/56&1/1&1/1 &0/3& 0/6& 18/32&& &23/154&30\\
6441&2/50 &0/28&0/1& 0/0&0/17&2/9&43/82&&0/10&47/187&21\\
6528&1/1&1/1&0/0&0/0&1/1&0/0&4/4&&&7/7&21
\\
6638&3/10&2/18&0/0&0/0&0/0&0/0&3/9&&0/25&8/37&21\\
6652&0/3&0/1&0/0&0/0&1/2&0/1&0/2&&1/5&1/9&21\\
6712& 0/10 & 0/4 & 0/0 & 0/0 & 2/2 &0/0& 5/11 &&0/8  &7/27 & 31\\
6779 (M56)& 0/1 &0/2&0/0&1/1 &3/3&0/2&0/3&&1/6&4/12&32\\
6934& 3/68 & 0/12 & 0/0 & 3/4 &0/0 & 2/3 &3/5 &  &1/6&11/92 &33\\
6981 (M72)&8/37 &3/7&0/0&3/3&0/0&0/0&0/1&& &14/48&12\\
7078 (M15)&0/65 &0/67&0/32&0/4&0/3&0/2&0/3&&0/11&0/176&13\\
7089 (M2)&5/23 &3/15&0/0&0/2&0/0&0/4&0/0&&0/12&8/44&14\\
7099 (M30)&1/4&2/2&0/0&2/2&1/6&0/0&0/0&&0/3&6/14&17\\
7492 &0/1&0/2&0/0&2/2&0/0&0/0&1/2&&&3/7&16\\

Pal 13 &0/4 & 0/0 &0/0 & 0/0 & 0/0& 0/0 & 1/1 && & 1/5 & 34\\
\hline
Total per type &41/713 &35/448 &6/54 &39/153 &16/140 &9/57 &157/316&0/5&23/161 &303/1886\\
\hline
\end{tabular}
\center{\quad $\dagger$. The variable star types are adopted from the General
Catalog of Variable Stars (Kazarovets et al. 2009; Samus et al. 2009). Entries
expressed as
M/N indicate the M variables found or reclassified by our program and the total number
N of presently known variables. Column 11 indicates the relevant papers on a given clusters.

\quad $*$. Numbers from this column are not considered in the totals. Here we include unclassified variables or likely field variables in the FoV of the cluster.

\quad References: 1. \citet{Arellano2013b}; 2. \citet{Kains2012}; 3.
Arellano Ferro et al. (2014a); 4.
Arellano Ferro et al. (2004); 5. \citet{Kains2015}; 6. Arellano Ferro et al.
(2011); 7. \citet{Bramich2012}; 8. Arellano Ferro et al. (2010); 9. Arellano Ferro
et al. (2008a), 10. Arellano
Ferro et al.
(2013a), 11. Arellano Ferro et al. (2008b), 12. \citet{Bramich2011}; 13. \citet{Arellano2006}; 14. L\'azaro et al. (2006); 15. \citet{Kains2013}; 
16. Figuera Jaimes et al. (2013); 17. \citet{Arellano2015a}, 18. Arellano
Ferro et al. (2016); 19. Arellano Ferro
et al. (2015b); 20. Tsapras et al. (2017); 21. \citet{Skottfelt2015}; 22. \citet{Arellano2019}; 23. \citet{Arellano2018a}; 24. \citet{Deras2018}; 25. \citet{Deras2019}; 26. \citet{Arellano2020}; 27. \citet{Yepez2020}; 28. \citet{Arellano2018b}; 29. \citet{Ahumada2021}; 30. \citet{Yepez2022}; 31. \citet{Deras2020}; 32. \citet{Deras2022}; 33. \citet{Yepez2018}; 34. \citet{Yepez2019}; 35. \citet{Lata2019}. }
\end{center}

\end{table*}

\section{Conclusions}

A homogeneous approach towards the determination of mean $M_V$ and [Fe/H] from the Fourier decomposition of the cluster member RRL light curves, enables a new empirical exploration of the nature of the $M_V$-[Fe/H] relation, which describes the dependence of the luminosity of the HB on the metallicity. Although numerous efforts, from assorted strategies, have been performed to establish the zero point and slope of the relation, universal values have been elusive. We found that, if the RRL stars are to be used as indicators of the form of the relation, or if this is to be employed as a distance indicator instrument, it should be treated
independently for RRab and RRc stars. The reason is that the relation displays a different nature; for the RRab stars it is non linear with considerable scatter, while for the RRc it is tight, linear and the slope is mild.

Following the suggestion of theoretical works \citep{Demarque2000}, the inclusion of the HB structural parameter $\mathcal L$ demonstrates that $M_V$ is also correlated with $\mathcal L$, in a nonlinear fashion for the RRab analysis. For the RRc the role of $\mathcal L$ is negligible. We offer a calibration $M_V$-[Fe/H]-$\mathcal L$, with [Fe/H] in the spectroscopic scale of \citet{Carretta2009} (eq. \ref{MVFELab}) or in the \citet{Nemec2013} scale (eq. \ref{MVFELnab}) valid for RRab stars, and linear calibrations $M_V$-[Fe/H] in the above two scales (eq. \ref{FEHUVc} or eq. \ref{FEHNc}) valid for RRc stars. 

We find pertinent at this point to recollect a well established result, that globular clusters harbour more that one generation of stars \citep[e.g.][]{Bedin2004,Piotto2005,Piotto2007}, and that each generation has a measurable different chemical abundance, particularly the He content increases in later generations \citep{Milone2018}. As a result of different evolutionary sequences, the objects on the HB have a large range in He abundance causing the observed HB structure, particularly the colour range, and the breadth; the higher the value of Y, the more luminous the corresponding ZAHB would be. Recently it has been shown that small variations in the  He-burning core mass would also contribute to the observed breadth of the HB \citep{Yepez2022}. Hence, the values of the $\mathcal L$ parameter employed in the present investigation may be responding to these effects, which in turn may be responsible, at least partially, of the scatter observed in the correlations. Therefore, while it is helpful identifying the cluster star members, as we have done, the identification of stars belonging to the different generations in each cluster might offer an important improvement on the calibration of the $M_V$-[Fe/H]-$\mathcal L$ correlation.

To give support to our results, we compared the Fourier determinations of [Fe/H] with the spectroscopic values in the scale of \citet{Carretta2009} and found them to be in excellent agreement.
The distances obtained from the mean Fourier $M_V$ calibrations, and their zero points, have proven to match within 1.7 kpc and displaying a difference dispersion with an rms of 0.7 kpc, with the independent and also homogeneous distances determined by \citet{Baumgardt2021} from $Gaia$-eDR3 and HST data.

The CCD time-series photometric study of globular clusters, in combination with
difference
image analysis, has been be very fruitful in the discovery of new
variables. Following the latest version of the Catalogue of Variable Stars in Globular Clusters \citep{Clement2001}, we updated in Table \ref{NoVARS} the number of know variables per type and per cluster and indicate the numbers of variables found and classified by our program over the years, for a total of 303 out of the 1886 variables, likely cluster members, presently known in the family of the 35 clusters considered. 

\vskip 1.0cm

\section{ACKNOWLEDGMENTS}
I am indebted with all my colleagues that have coauthored the numerous papers cited here, for their sincere dedication and implementation of their expertise to the several astrophysical fronts involved in the project. My special thanks to Prof. Sunetra Giridhar and Dr. Dan Bramich for extended discussions and very useful suggestions. A warm acknowledgement to Dr. Ivan Bustos Fierro for his willingness to decipher the stellar membership status in several clusters. 
I recognize with gratitude the many corrections and comments made by a very attentive anonymous referee, which triggered relevant modifications in the manuscripts. I am grateful to the institutions operating the employed telescopes for the time granted to our project, and to the many support staff members in all these the observatories for making possible and efficient all our data gathering. The facilities at IAO and CREST are operated by the Indian Institute of Astrophysics, Bangalore. The project has been generously supported through the years by the program PAPIIT of the DGAPA-UNAM, M\'exico via several grants, the most recent being IG100620. 

\section*{DATA AVAILABILITY}
The data underlying this article shall be made available on request to the author (armando@astro.unam.mx).

\bibliographystyle{rmaa}
\bibliography{MvFeH} 

\end{document}